Stability and electronic properties of layered NaMnO$_2$ using the SCAN(+*U*)

Haeyoon Jung[1], Jiyeon Kim[1,2*], Sooran Kim[1*]

[1]Department of Physics Education, Kyungpook National University, Daegu 41566, South Korea

[2]The Center for High Energy Physics, Kyungpook National University, Daegu 41566, South Korea

*Corresponding authors: mygromit@gmail.com, sooran@knu.ac.kr

## Abstract

Considering electron correlation appropriately is important to predict the properties of layered transition metal oxides, which have drawn a lot of attention as cathode materials for sodium-ion batteries. Here, we explore the phonon and electronic properties of layered $NaMnO_2$ using the recently developed strongly constrained and appropriately normed (SCAN) functional. We also introduce the Coulomb interaction $U$ to find an accurate description of Mn $3d$ orbitals. The phonon dispersion curves show the structural stability with the SCAN, which is consistent with prior experimental stability at high Na concentrations. On the other hand, imaginary phonon frequencies were observed by applying $U$, which indicates structural instability. Namely, SCAN properly describes the phonon properties of layered $NaMnO_2$ whereas SCAN+$U$ does not. We further explore the Jahn-Teller (J-T) stability and magnitude of J-T distortion depending on $U$ and find that SCAN results are consistent with Perdew-Burke-Ernzerhof functional, PBE+$U$ results. Our results suggest that SCAN itself properly describes the physical properties of $NaMnO_2$ without adding $U$ explicitly.

## Keywords

# 1. Introduction

Rechargeable batteries have attracted a lot of interest as the demand for advanced energy storage technology is rapidly increasing around the world [1]. Lithium-ion battery (LIB) is one of the most successful rechargeable batteries and has already been commercialized in many places. However, the uneven distribution of lithium and increase of production cost [2,3] require an alternative energy source, which can replace LIBs. On the other hand, sodium is the second lightest alkali metal with abundant reserves and is readily available worldwide. Because of these advantages, sodium-ion battery (SIB) has been thought of as a promising alternative to LIB [3,4].

Among the cathode materials for SIBs, layered $NaMnO_2$ has been intensively studied for large-scale battery applications because of its safeness and abundance [5–15]. The $NaMnO_2$ crystallizes in a monoclinic unit cell with the space group *C*2/*m* and exhibits the Jahn-Teller (J-T) distortion from the $Mn^{3+}$ ion. This crystal structure is a classified O3 type where sodium ions occupy the octahedral sites. The number 3 denotes the number of three different $MnO_6$ layers (AB, BC, and CA). The J-T distorted structure of the O3-type is represented as O'3. Figure 1 shows the crystal structure of O'3-$NaMnO_2$ and the energy level of $Mn^{3+}$ *d* orbitals. The J-T distortion splits the energy level of the degenerate *d* orbital and lowers the energy of the structure. O'3-$NaMnO_2$ (distorted structures of O3-type) is thermodynamically stable at the high contents of Na [5,6,16]. A high discharge capacity of 197 mAh/g was reported for O'3-$NaMnO_2$, but the capacity decayed after 20 cycles without significant structural change [5]. The cyclability was improved by doping, for example, $Na(Mn_{0.89}Cu_{0.08}Sb_{0.03})O_2$ exhibits 204 mAh/g discharge capacity and capacity retention of 80 % after 150 cycles [8]. Furthermore, Na ion ordering and Mn charge ordering of $Na_{5/8}MnO_2$ with the cooperative Jahn-Teller distortion develop the magnetic ordering at low temperature, which shows a strong interaction among Na ion, charge, and magnetic orderings in manganese oxides [14].

Since $NaMnO_2$ is one of the transition metal oxides, considering strong electron-electron correlation is essential for the accurate description of the physical properties of $NaMnO_2$. The use of tunable Hubbard $U$ parameters within DFT+$U$ formalism greatly improves the reproducibility of calculations for experimental physical properties for layered transition metal cathode materials [17–23]. Thus, many previous theoretical studies have been done to explain the properties of $NaMnO_2$ with the inclusion of parametric Coulomb interaction $U$ in a range of 2.5 to 5.0 eV under generalized gradient approximation (GGA) or Perdew-Burke-Ernzerhof (PBE) functional [24–29].

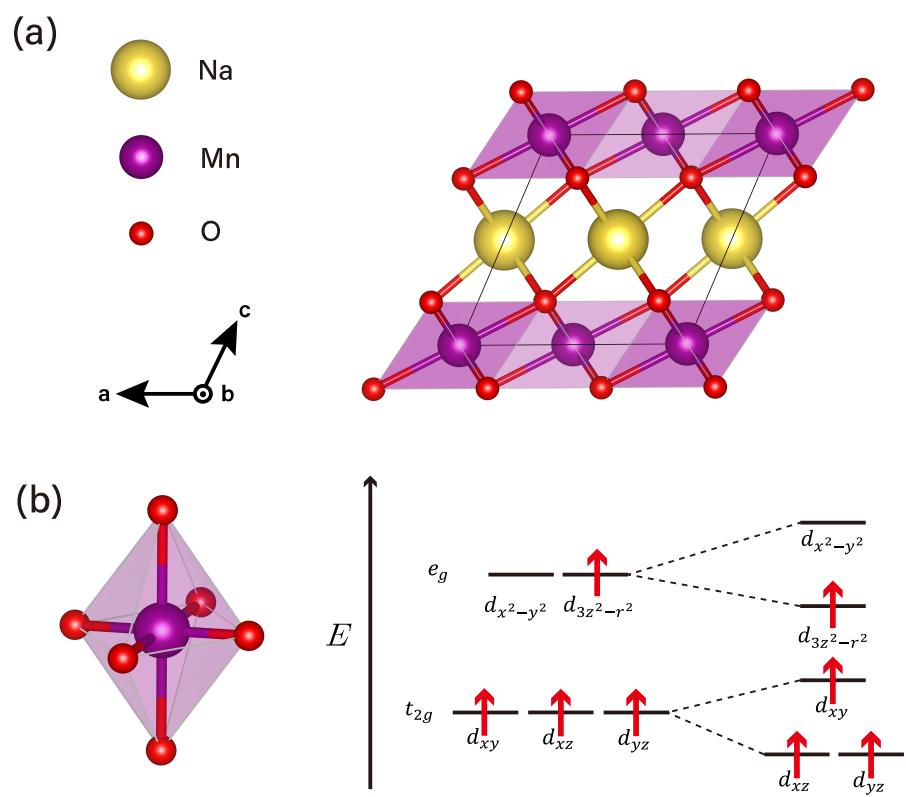

Fig. 1 (a) Crystal structure of O'3-$NaMnO_2$ (b) Octahedral structure of $MnO_6$ with the J-T distortion and electronic configurations in the $d$-orbital energy level of $Mn^{3+}$ ion

Recently, strongly constrained and appropriately normed (SCAN) was developed, a non-empirical meta-GGA functional [30,31]. This functional does not require external tunable parameter $U$ to describe the properties of many transition metal oxides [32–34]. However, previous studies also reported that SCAN functional still exhibits a self-interaction error [19,35–39], which requires the Coulomb interaction $U$ to reduce the error. Therefore, it is worth investigating the utility of the SCAN functional for each material.

In this paper, we have investigated the physical properties of layered O'3-NaMnO$_2$ using the recently developed SCAN functional, especially focusing on phonon and lattice dynamics. We also considered the Coulomb interaction $U$ to examine the accuracy of SCAN functional for NaMnO$_2$. The phonon dispersion curves with the SCAN show the structural stability without imaginary phonon frequencies while the phonon soft modes are observed with the SCAN+$U$. Because O'3-NaMnO$_2$ is an experimentally synthesized stable structure, the phonon calculations clearly present that SCAN describes the properties of O'3-NaMnO$_2$ better than SCAN+$U$. The calculated lattice parameters within the SCAN(+$U$) approach are compared with the experimental lattice parameters. Furthermore, we explore the J-T stability and the magnitude of J-T distortion under different $U$ values and compare SCAN(+$U$) results with PBE+$U$ calculations.

## 2. Computational method

All spin-polarized density functional theory (DFT) calculations were performed by the Vienna *ab initio* simulation package (VASP), which implements a pseudo potential band approach [40,41]. We utilized SCAN (strongly constrained and appropriately normed semilocal density functional) [30,31] as an exchange-correlation functional and also employed Perdew-Burke-Ernzerhof (PBE) functional for comparison [42]. We further included Coulomb correlation $U$ to account for the correlated $d$ orbitals of the Mn atom and used the Dudarev method for the double-counting correction [43]. The effective on-site correlation, $U_{eff} = U - J$ of 3.9 eV is chosen unless otherwise specified [44]. To investigate the dependence on $U$ in SCAN+$U$ approach, the various $U_{eff}$ values were used (0, 1, 2, 3, 3.24, 3.62, and 3.9 eV). $U_{eff}$ of 3.24 eV and 3.62 eV are from the previous constrained random phase approximation calculations for layered $NaMnO_2$ [21]. The energy cutoff for the plane waves is 520 eV. The k-point sampling is 6 × 13 × 6, which corresponds to the *k*-point density of 5000/atom. The monoclinic $NaMnO_2$ with the space group *C*2/*m* has the experimental lattice parameters: $a$ = 5.673 Å, $b$ = 2.856 Å, $c$ = 5.807 Å, and $β$ = 113.290° [16]. We fully relaxed the lattice parameters and the atomic coordinates from the initial experimental structure until the force was less than 0.01 eV/Å.

We employed PHONOPY to calculate phonon dispersion and density of states [45,46]. The force constants and dynamic matrix were calculated from supercells with finite displacements based on the Hellmann-Feynman theorem. The 2 × 2 × 2 supercell and the 3 × 7 × 3 *k*-point sampling were used for the phonon calculations.

# 3. Results and discussions

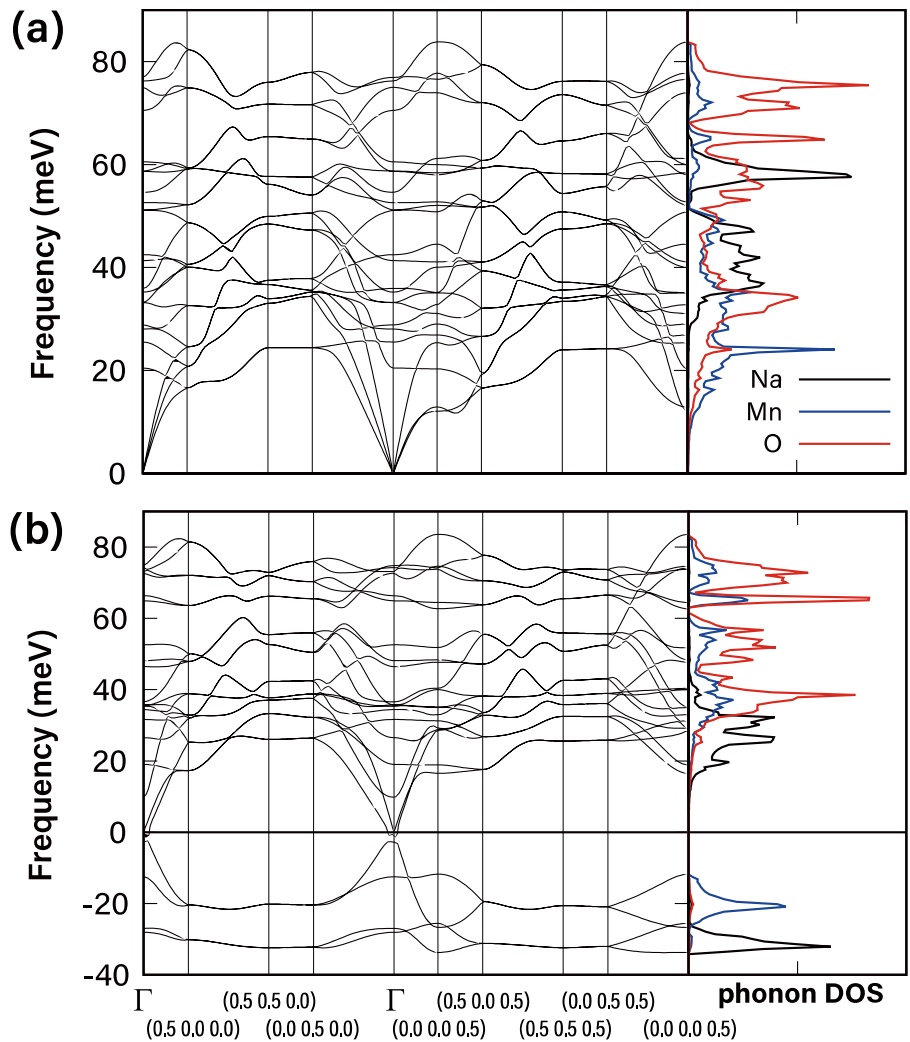

Fig. 2 Phonon dispersion curves and phonon density of states of O'3-NaMnO$_2$ with (a) the SCAN (b) the SCAN+$U$. The imaginary phonon frequencies infer the instability of the structure.

Figure 2 displays the phonon structures of layered O'3-NaMnO$_2$ using the SCAN and SCAN+$U$ methods. In our calculations, O'3-NaMnO$_2$ consists of a total of 8 atoms: 2 Na, 2 Mn, and 4 O atoms. Therefore, 24 modes are generated from three degrees of freedom. Among them, three acoustic modes with zero energy at the Γ point and 21 optical modes can be observed in the phonon bands.

The phonon bands with SCAN (Fig. 2 (a)) show that O'3-NaMnO$_2$ is dynamically stable since imaginary phonon frequencies, i.e. soft phonon modes, are not observed in the phonon structure. This stable phonon dispersion is consistent with the experimental result of stable O'3-NaMnO$_2$ [5,16]. From the phonon DOS in Fig. 2 (a), O atoms contribute mostly in the high-frequency region above about 50 meV, while the displacements of Mn and Na ions dominate in the low and medium region in frequency because of their atomic mass. This vibration distribution is similar to the phonon of O'3- NaMnO$_2$ with the PBE+$U$ [24] and LiMnO$_2$ [47].

On the other hand, the phonon structure with SCAN+$U$ in Fig. 2 (b) exhibits imaginary phonon frequencies through all high symmetry points. This phonon dispersion indicates that O'3-NaMnO$_2$ is not stable in contrast to the phonon bands with the SCAN and the previous experiments. The soft phonon modes mostly originate from the lattice vibrations of Na and Mn ions as shown in the phonon DOS in Fig. 2 (b). Namely, the phonon bands with SCAN properly reproduce the experimental stability of O'3-NaMnO$_2$ whereas those with the SCAN+$U$ does not. It suggests that SCAN+$U$ is not appropriate to describe the properties of O'3-NaMnO$_2$.

Table 1. Lattice parameters of O'3-NaMnO$_2$. The relative errors of calculated values are in parenthesis.

| Lattice parameters | SCAN | SCAN+$U$ ($U$=3.9 eV) | exp. [16] |
| --- | --- | --- | --- |
| $a$(Å) | 5.695 (0.40%) | 5.618 (-0.96%) | 5.673 |
| $b$(Å) | 2.859 (0.10%) | 2.896 (1.42%) | 2.856 |
| $c$(Å) | 5.754 (-0.91%) | 5.734 (-1.26%) | 5.807 |
| $β$(°) | 113.4 (0.17%) | 112.5 (-0.65%) | 113.3 |

We have compared the experimental [16] and calculated lattice parameters using different $U$ values. Table 1 shows the structural information of O'3-NaMnO$_2$ and the relative errors of each calculated value compared to the experimental one. The lattice parameters with the SCAN have a relative error of less than 1% compared to the experimental values. It is worth noting that the calculated values with the SCAN have smaller relative errors than those of SCAN+$U$ in all parameters. As $U$ increases, the lattice parameter $a$ and $c$ tend to decrease while the lattice parameter $b$ increases. The result that the SCAN predicts the lattice parameters better than SCAN+$U$ suggests that the SCAN approach would be more appropriate than SCAN+$U$ for O'3-NaMnO$_2$, which is consistent with our phonon calculations.

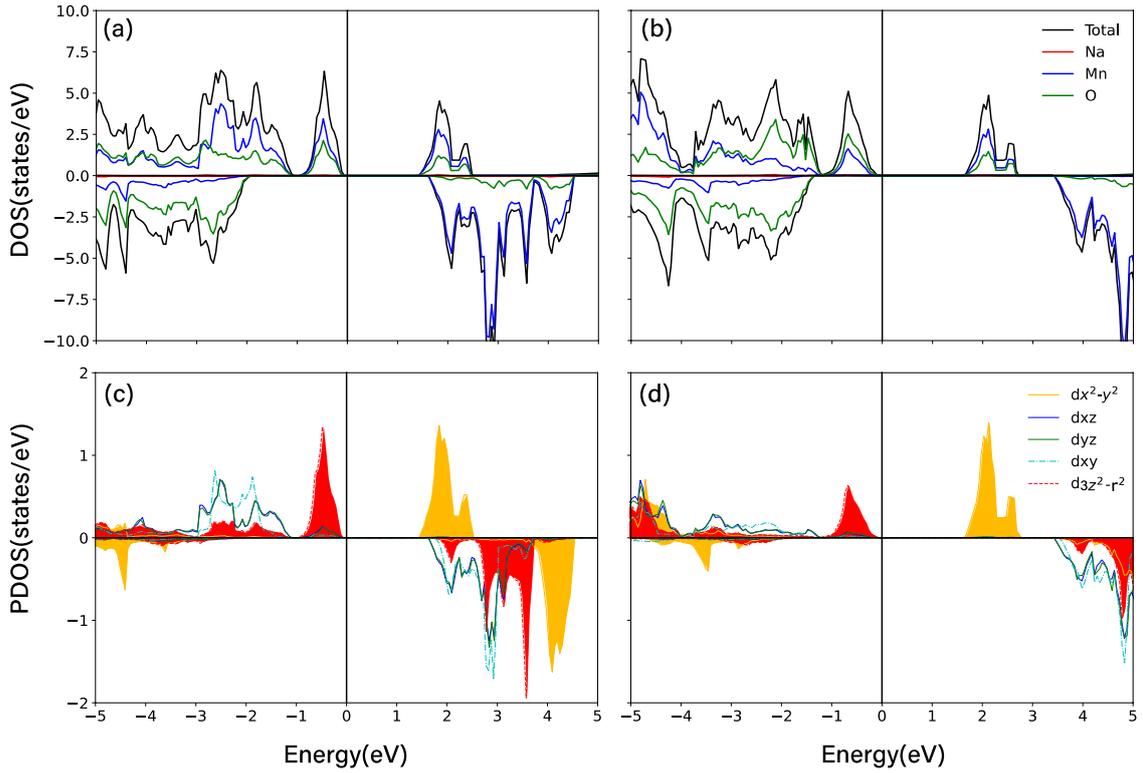

Fig. 3 Total and atomic density of states (DOS) with (a) the SCAN and (b) the SCAN+$U$. Projected density of states (PDOS) of Mn $d$ orbitals with (c) the SCAN and (d) the SCAN+$U$. The PDOS curves of $d_{x^2-y^2}$ and $d_{3z^2-r^2}$ orbitals are filled with yellow and red colors, respectively.

We further investigate the electronic DOS of O'3-NaMnO$_2$ within the SCAN and SCAN+$U$ approach as shown in Fig. 3. The Mn$^{3+}$ ions in O'3-NaMnO$_2$ are in the high spin states with $3d^4$ electrons. The calculated magnetic moment values of Mn ion from the SCAN and SCAN+$U$ are 3.676 $\mu_B$ and 3.892 $\mu_B$, respectively. The slight increase of the magnetic moment with $U$ would be from the localization of Mn-$d$ by applying $U$. The band gap obtained from SCAN is 1.45 eV, which is similar to the band gap of ~1.3 eV from the previous PBE+$U$ ($U$ = 3.9 eV) calculations [24,28]. The band gap increases to 1.65 eV with the inclusion of the Coulomb correlation, $U$. In addition, the down spin states increase in energy with $U$: the oxygen valence bands and the Mn conduction bands are shifted up from about -2 eV to -1eV and from about 1.5 eV to 3.5 eV, respectively.

Figure 3 (c) and (d) show the Mn $d$ orbital decomposed DOS of O'3-NaMnO$_2$ with the SCAN and SCAN+$U$, respectively. Both DOSs exhibit the clear splitting of $d_{3z^2-r^2}$ and $d_{x^2-y^2}$, which results in the opening of the band gap. This splitting is obtained by the J-T distortion of the elongated Mn-O bonds aligned in the octahedral local z-direction. The inclusion of $U$ with SCAN functional does not much alter the shape of the orbitals near the Fermi level, but shifts the energy of orbitals, in particular, by increasing the down spin states in conduction bands and decreasing $t_{2g}$ orbitals in valence bands.

We have analyzed the energy and the magnitude of the J-T distortion using different $U$ values in SCAN(+$U$). Figure 4 (a) shows the energy difference $\Delta E = E_{JT} - E_0$ between the structures with and without the J-T distortion where $E_{JT}$ and $E_0$ indicate the energy with and without the J-T distortion respectively. The total energy of the J-T distorted structure is lower than that of the undistorted structures regardless of $U$ values. The energy difference $\Delta E$ from the J-T distortion is -0.372 eV at $U$ = 0 eV and slightly becomes larger up to -0.434 eV at $U$ = 3.9 eV. We have compared the $\Delta E$ using the SCAN(+$U$) with that using the PBE+$U$ ($U$ = 3.9 eV), indicated as a dotted line in Fig. 4 (a). The $\Delta E$ difference between PBE+$U$ and SCAN(+$U$) becomes larger as $U$ increases: The $\Delta E$ differences of two functionals are 0.017 eV and 0.079 eV for SCAN and SCAN+$U$ ($U$ = 3.9 eV). Namely, the results with the SCAN itself exhibit the closest results to that with the PBE+$U$ ($U$ = 3.9 eV). This result is consistent with the prior studies that SCAN in general does not require explicit correction of $U$ or at least requires smaller $U$ values than PBE functional to describe the properties of materials [19,32–34,37–39].

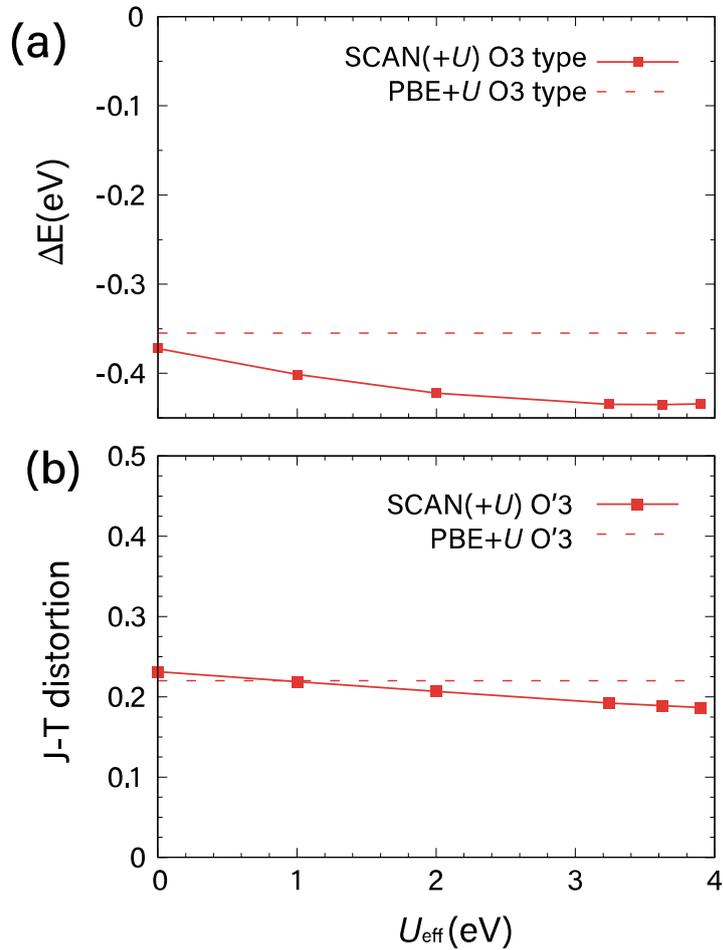

Fig. 4 $U$ dependence of (a) the energy difference of J-T distorted and undistorted NaMnO$_2$ and (b) magnitude of the J-T distortion. The dotted lines indicate the results with the PBE+$U$ ($U$ = 3.9 eV).

Furthermore, we evaluated the magnitude of the J-T distortion as a function of the $U$ value and compared it with the PBE+$U$ as shown in Fig. 4 (b). The J-T distortion with the elongation along the local $z$-axis of MnO$_6$ consists of two long bonds and four short bonds of Mn-O. The magnitude of the J-T distortion is defined as in

$$\text{J-T distortion} = \frac{6(l_{long} - l_{short})}{2l_{long} + 4l_{short}} \qquad (1)$$

where $l_{long}$ and $l_{short}$ indicate the long and short bond lengths of Mn-O, respectively [48]. For example, with the SCAN, the long and short bond lengths are 2.410 Å and 1.927 Å, respectively, which gives the J-T distortion magnitude of 0.231. The J-T distortion slightly decreases as increasing of the $U_{eff}$, whose trend is consistent with that using the PBE functional [24]. SCAN(+$U$) ($U$ = 0-1eV) gives the closest J-T distortion with the PBE+$U$ ($U$ = 3.9 eV). This result also suggests that the SCAN functional itself well describes the strong electron-electron interaction of NaMnO$_2$ without considering significant explicit $U$ inclusion as in the PBE functional.

## 4. Conclusion

In conclusion, we have explored the lattice dynamics and electronic properties of O'3-NaMnO$_2$ using the SCAN(+$U$) approach and demonstrate that SCAN without the direct inclusion of the Coulomb interaction $U$ properly reproduces the experimental stability and lattice parameters. The phonon dispersion curve with the SCAN exhibits the dynamical stability of O'3-NaMnO$_2$, which is consistent with previous experimental observation. On the other hand, the phonon band with the SCAN+$U$ shows the phonon soft modes implying structural instability. This phonon result suggests that the SCAN functional itself is an appropriate method to describe O'3-NaMnO$_2$. The calculated lattice parameters with the SCAN agree well with the experimental lattice parameters. Furthermore, we found that SCAN with smaller $U$ ($U$ < 1eV) gives similar results with the PBE+$U$ ($U$ = 3.9 eV) according to the energy difference $\Delta E$ and the magnitude of the J-T distortion, which shows the SCAN functional describes electron correlation of O'3-NaMnO$_2$ more appropriately over PBE functional without sizable inclusion of $U$. This work demonstrates that lattice properties, especially phonon structures would be useful to find the appropriate computational approach for materials. We hope that this study can promote further investigation of the performance of SCAN functional for Na-ion cathode materials.

## Acknowledgements

This work was supported by the Korea Electric Power Corporation (Grant R20XO02-12) and KISTI Supercomputing Center (Project No. KSC-2021-CRE-0495).